# "Zero Trust Chain: A Design Pattern for Improved Interoperability and Security in Polkadot"

*Santiago Márquez Solís*
*(April 2023 – Madrid)*

*Abstract*:

*This research article presents various design patterns for improving interoperability in Polkadot, a blockchain platform. These patterns include chain bridges, interoperability standards, common asset identifiers, governance agreements, oracle chains, and a hypothetical design pattern called Zero Trust Chain. Implementation of these design patterns can help improve security and confidence in transactions between different chains on the Polkadot network, allowing for faster and more efficient communication. The article also emphasizes the importance of interoperability in blockchain technology and highlights Polkadot's flexibility in creating customized specialized chains that can further improve interoperability on the network. Overall, this article highlights how design patterns can improve interoperability in Polkadot, which could lead to greater adoption of blockchain technology in various industries.*

## 1. Introduction. Context to the Interoperability Problem

Interoperability refers to the ability of different systems, platforms, or blockchains to communicate and work together effectively. In the context of blockchain technology, interoperability refers to the ability of different blockchains to interact with each other and allow for the transfer of data and assets from one chain to another. Interoperability is crucial for the widespread adoption of blockchain technology, as it allows different blockchains to communicate and transact more efficiently and securely. Without interoperability, blockchains would exist in separate silos, limiting their ability to offer services and benefits to a wide range of users and applications.

Interoperability has been a challenge in traditional networks, as different networks and systems often use different protocols, languages, and technologies that are incompatible with each other. This has led to the development of intermediary technologies such as gateways, proxies, and middleware to enable communication between different systems. However, these solutions are often complex, expensive, and can create additional points of failure and security vulnerabilities. In recent years, efforts have been made to standardize protocols and technologies to improve interoperability in traditional networks, but the challenge remains significant. In contrast, blockchain technology is inherently designed for interoperability, with the potential to enable seamless communication and transactions between different chains and networks.

Interoperability is a fundamental aspect of blockchain technology that plays a critical role in enabling the full potential of decentralized applications and services. In this context, interoperability refers to the ability of different blockchains to interact and communicate with each other seamlessly, allowing for the transfer of data and value across different networks. Here are three paragraphs explaining why interoperability is important in blockchain technology:

- First, interoperability enables the creation of decentralized ecosystems that can offer a wide range of services and applications to users. This is because it allows different blockchains to connect and work together, allowing developers to build complex applications that leverage the unique features of different chains. For example, a decentralized finance (DeFi) application may need to access data from multiple blockchains to perform its operations, and interoperability enables this by allowing for seamless communication between different chains. Interoperability also allows for the creation of cross-chain smart contracts that can execute transactions and events across different chains, unlocking new possibilities for decentralized applications.
- Second, interoperability is essential for achieving scalability in blockchain networks. As the number of users and applications on blockchain networks grows, the demand for transaction processing and data storage increases. Interoperability allows different chains to share the load of transaction processing and data storage, enabling the creation of scalable and efficient blockchain networks. Interoperability also reduces the need for complex layer-2 solutions that are required to scale individual chains, further improving the efficiency and scalability of blockchain networks.
- Third, interoperability is important for fostering innovation and competition in the blockchain industry. Without interoperability, different blockchains would exist in separate silos, limiting their ability to offer unique features and services to users. Interoperability allows different chains to collaborate and compete with each other, fostering innovation and driving the development of new applications and services. This benefits users by providing them with a wider range of options and features to choose from, and it benefits developers by enabling them to build more powerful and flexible applications.

In conclusion, interoperability is a crucial aspect of blockchain technology that enables the creation of decentralized ecosystems, improves scalability, and fosters innovation and competition. As the blockchain industry continues to grow and evolve, interoperability will become even more important, allowing for the development of new applications and services that can unlock the full potential of decentralized technology.

## 1.1. Article Structure

The structure to address the objectives has been articulated based on the following points:

- **Point 2: State of the Art of Interoperability Solutions in Polkadot**: Brief description of the Relay Chain and Parachains structure in Polkadot and the description of current interoperability solutions in Polkadot, including chain bridges, interoperability standards, common asset identifiers, governance agreements, and oracle chains.
- **Point 3: Zero-Knowledge Proofs: What They Are and How They Work**: Definition and explanation of zero-knowledge proofs (ZKPs) and description of different types of ZKPs and their applications in blockchain technology. We also analyzed examples of blockchain projects that use ZKPs to improve the privacy and security of transactions.
- **Point 4: Proposal for Interoperability in Polkadot through the Use of Zero-Knowledge Proofs**: Explanation of how ZKPs could be applied to improve interoperability in Polkadot, including the proposal of a hypothetical design pattern called "Zero Trust Chain." and detailed description of how the ZKP proposal would be implemented in the Polkadot network and how it would improve interoperability between chains.
- **Point 5: Conclusions**: Summary of the main findings and conclusions of the article and the discussion of the implications of the ZKP proposal for improving interoperability in Polkadot and its possible impact on the adoption of blockchain technology. Limitations and possible areas for future research in the field of blockchain interoperability.

## 2. State of the Art of Interoperability Solution in Polkadot.

Interoperability is a critical aspect of blockchain technology that enables different chains to communicate and interact with each other seamlessly. In the context of Polkadot, interoperability is essential for achieving the full potential of its Relay Chain and Parachains structure, which enables different chains to connect and work together. However, achieving interoperability between different chains is a complex challenge that requires innovative solutions. In this section, we will review the current state of the art of interoperability solutions in Polkadot, including chain bridges, interoperability standards, common asset identifiers, governance agreements, and oracle chains. We will also discuss the limitations and challenges of current interoperability solutions and the importance of improving interoperability in Polkadot.

Relay Chains and Parachains are two important components of the Polkadot network. The Relay Chain is the main chain that provides the basic infrastructure for the entire network. It manages the shared security, consensus, and governance of the network and facilitates communication and coordination between different chains. Parachains, on the other hand, are independent chains that run in parallel to the Relay Chain. They are designed to be highly customizable, enabling developers to create specialized blockchains with unique features and functionalities.

To connect to the Polkadot network, a chain must become a Parachain by leasing a slot on the Relay Chain. Once a chain becomes a Parachain, it gains access to the shared security and other resources of the network, and it can communicate and interact with other chains in the network. Parachains can also interact with each other through cross-chain messaging.

The Relay Chain manages the shared security and consensus of the network through a unique consensus mechanism called Nominated Proof-of-Stake (NPoS). NPoS is a variation of Proof-of-Stake (PoS) that allows token holders to nominate validators to secure the network. Validators are responsible for producing new blocks on the Relay Chain and verifying transactions on Parachains. The Relay Chain also has a set of governance mechanisms that allow token holders to vote on network upgrades, parameter changes, and other decisions.

## 2.1. Chain Bridges

Chain bridges are one of the current interoperability solutions in Polkadot that allow communication and transfer of value between different chains. Chain bridges are essentially software modules that enable two chains to communicate with each other by facilitating the transfer of assets or data from one chain to another.

Chain bridges typically work by locking the assets on one chain and minting corresponding assets on the other chain. For example, a chain bridge between Polkadot and Ethereum may lock a certain amount of ETH on the Ethereum network and mint an equivalent amount of a Polkadot-compatible token on the Polkadot network. This token can then be used on the Polkadot network or transferred back to the Ethereum network by burning the tokens on Polkadot and unlocking the corresponding ETH on Ethereum.

One example of a chain bridge in Polkadot is the Bifrost bridge, which allows interoperability between Polkadot and the Binance Smart Chain (BSC). The Bifrost bridge enables the transfer of tokens and other assets between the two networks, as well as the execution of cross-chain smart contracts. Another example is the Darwinia bridge, which enables interoperability between Polkadot and Ethereum. The Darwinia bridge supports cross-chain token transfers, and it also allows for the creation of Ethereum-compatible smart contracts on the Polkadot network.

While chain bridges provide a way to achieve interoperability between different chains, they have some limitations. For example, they require the cooperation of both chains, and they can be complex and costly to implement. Additionally, they may introduce security risks and centralization concerns, as the bridging process often involves trusted intermediaries.

## 2.2. Interoperability Standards

Interoperability standards are another current interoperability solution in Polkadot that aim to establish common rules and protocols for communication and interaction between different chains. Interoperability standards can help to ensure compatibility and seamless integration between different chains, and they can also simplify the development and deployment of cross-chain applications.

In Polkadot, one example of an interoperability standard is the Cross-Chain Message Passing (XCMP) protocol. XCMP is a protocol that allows different parachains to communicate and exchange messages with each other, even if they are built using different technologies or programming languages. XCMP enables different parachains to share data, assets, or execute cross-chain transactions, and it provides a secure and reliable way to achieve interoperability within the Polkadot network.

An example of an interoperability standard in Polkadot is the Polkadot Interoperability Protocol (PIP). PIP is a proposed standard that aims to establish a common set of rules and protocols for interoperability between Polkadot and other blockchain networks. PIP defines a set of APIs and data formats that allow different chains to communicate and exchange data in a standardized and secure way.

Interoperability standards can improve the interoperability between different chains, but they have some limitations as well. For example, interoperability standards may require a significant effort to design and implement, and they may not be universally adopted by all chains. Additionally, interoperability standards may not be sufficient to achieve full interoperability between different chains, as they may not be able to address all the technical and economic challenges of cross-chain communication and coordination.

## 2.3. Common Assets Identifiers

Common asset identifiers are another current interoperability solution in Polkadot that aim to establish a common language for identifying and tracking assets across different chains. Common asset identifiers can help to ensure compatibility and seamless integration between different chains, and they can also simplify the development and deployment of cross-chain applications.

In Polkadot, one example of a common asset identifier is the Polkadot Asset Specification (PAS). PAS is a proposed standard that defines a common format for representing and identifying assets on the Polkadot network. PAS defines a set of metadata that provides information about the asset, such as its name, symbol, decimal precision, and total supply. By using a common asset identifier like PAS, different chains can easily identify and track the same asset across different chains, facilitating cross-chain interoperability.

For example of a common asset identifier in Polkadot is the Ethereum Address Format (EAF). EAF is a format for representing Ethereum addresses on the Polkadot network. By using EAF, Polkadot can interact with Ethereum-based assets and smart contracts using the same address format, facilitating cross-chain interoperability between Polkadot and Ethereum.

Common asset identifiers can improve the interoperability between different chains by establishing a common language for identifying and tracking assets. However, they have some limitations as well. For example, common asset identifiers may not be able to address all the technical and economic challenges of cross-chain communication and coordination, and they may not be universally adopted by all chains.

## 2.4. Oracle Chains

Oracle chains are another current interoperability solution in Polkadot that aim to provide reliable and secure data feeds to different chains. Oracle chains act as intermediaries between different chains and external data sources, enabling the secure transfer of data and information across different chains.

In Polkadot, one example of an oracle chain is the Chainlink Oracle. The Chainlink Oracle is a decentralized oracle network that provides reliable and secure data feeds to different chains. The Chainlink Oracle aggregates data from different sources and delivers it to the requesting chain in a secure and tamper-proof way. This enables different chains to access and use external data, such as market prices, weather data, or other real-world events, to execute smart contracts and other decentralized applications.

Another example of an oracle chain in Polkadot is the Polkadex Oracle. The Polkadex Oracle is an oracle network that provides reliable and secure data feeds to the Polkadex decentralized exchange. The Polkadex Oracle aggregates data from different sources, such as centralized exchanges or price aggregators, and delivers it to the Polkadex network in a secure and verifiable way. This enables the Polkadex network to access real-time market data and execute trades on behalf of users.

Oracle chains can improve the interoperability between different chains by providing reliable and secure data feeds. However, they have some limitations as well. For example, oracle chains may introduce security risks and centralization concerns, as the oracles act as trusted intermediaries between different chains and external data sources. Additionally, oracle chains may not be able to address all the technical and economic challenges of cross-chain communication and coordination.

## 2.5. Limitations and Challenges of current interoperability Solutions

While there are several interoperability solutions in Polkadot, they still face some limitations and challenges. One of the main challenges is achieving full interoperability between different chains. While solutions like chain bridges, interoperability standards, common asset identifiers, governance agreements, and oracle chains can improve the communication and interaction between different chains, they may not be able to address all the technical and economic challenges of cross-chain interoperability. For example, achieving consensus across different chains, managing transaction fees and other economic incentives, and ensuring the security and reliability of cross-chain transactions can be complex and challenging.

Another challenge is achieving sufficient adoption of interoperability solutions by different chains. Interoperability solutions require the cooperation and participation of different chains, and they may not be universally adopted by all chains. This can lead to fragmentation and inefficiencies in the cross-chain communication and interaction, reducing the overall benefits of interoperability.

Additionally, interoperability solutions may introduce new security risks and centralization concerns. For example, chain bridges and oracle chains may act as trusted intermediaries between different chains and external data sources, increasing the risk of attacks and manipulation. Governance agreements may also introduce centralization concerns, as they require a coordinated decision-making process that may be dominated by a small group of stakeholders.

In summary, while current interoperability solutions in Polkadot provide a promising foundation for achieving cross-chain communication and interaction, they still face several challenges and limitations. Addressing these challenges will require continued innovation and collaboration between different chains and stakeholders in the Polkadot ecosystem.

## 3. Zero-Knowledge Proofs: What They Are and How They Work

Zero-knowledge proofs (ZKPs) are a cryptographic technique that allows one party (the prover) to prove to another party (the verifier) that they possess certain information or knowledge, without revealing the information itself. In other words, ZKPs allow the prover to demonstrate knowledge of a secret or private information without actually revealing that information to the verifier.

A zero-knowledge proof is a type of cryptographic proof that allows one party to prove to another party that they know a certain secret information, without revealing what that information is.

To illustrate this concept, let's consider the following scenario: You have two pockets in your pants, one with three coins and another with five coins. You want to prove to a friend that you know which pocket has more coins, but you don't want to reveal the actual number of coins in either pocket.

To prove your knowledge without revealing the actual number of coins, you could use a zero-knowledge proof. Here's how it could work:

- Your friend would randomly select one of the pockets, without showing you which one it is.
- You would then add a secret number to the number of coins in the selected pocket. For example, let's say your secret number is 2 and your friend selected the pocket with 3 coins. You would add 2 to 3, for a total of 5.
- You would then reach into both pockets and show your friend that you have a total of 8 coins. Your friend would be convinced that you know which pocket has more coins, without learning the actual number of coins in either pocket.

This is a basic example of how zero-knowledge proofs can be used to prove knowledge of a secret without revealing the secret itself. In more complex applications, zero-knowledge proofs can be used to prove the validity of a transaction or to verify a user's identity without revealing sensitive information.

ZKPs have numerous applications in cryptography and beyond, including authentication, privacy-preserving data sharing, and digital signatures. The technique is based on complex mathematical algorithms that enable the prover to convince the verifier of their knowledge, without revealing any additional information about the knowledge.

The concept of zero-knowledge proofs was first introduced in the 1980s, and since then, significant advancements have been made in the field, making it a powerful tool in the world of cryptography and cybersecurity. By utilizing ZKPs, parties can verify their identity, prove ownership of a digital asset, or validate transactions on a blockchain network, all while keeping their sensitive information private and secure.

There are several types of zero-knowledge proofs (ZKPs), each with its own unique strengths and applications in blockchain technology. Some of the most commonly used types of ZKPs in blockchain include:

- **zk-SNARKs (Zero-Knowledge Succinct Non-Interactive Arguments of Knowledge)**: This type of ZKP is widely used in privacy-focused blockchains such as Zcash and is known for its high level of efficiency. zk-SNARKs are used to prove that a transaction is valid without revealing any additional information about the transaction, making them ideal for use in privacy-preserving blockchains.
- **Bulletproofs:** This type of ZKP is a more recent development and is used to prove the validity of range proofs, which are used to ensure that a value falls within a certain range without revealing the actual value. Bulletproofs are particularly useful in blockchain applications that require efficient verification of transactions, such as in the Lightning Network.
- **zk-STARKs (Zero-Knowledge Scalable Transparent Argument of Knowledge)**: This type of ZKP is a newer development in the field and is designed to be more transparent and auditable than zk-SNARKs. zk-STARKs are particularly useful in decentralized applications where transparency and auditability are important, such as in supply chain management or voting systems.
- **ZK-Proofs**: This type of ZKP is a more general category that encompasses a wide range of different proof systems. ZK-Proofs are particularly useful in applications that require a more flexible approach to proving knowledge, such as in identity verification or authentication systems.

There are several blockchain projects that utilize zero-knowledge proofs (ZKPs) to enhance the privacy and security of transactions. Here are some examples:

- **Zcash**: Zcash is a privacy-focused cryptocurrency that uses zk-SNARKs to enable private transactions. With ZKPs, Zcash users can conduct transactions without revealing any details about the sender, receiver, or transaction amount.
- **Monero**: Monero is another privacy-focused cryptocurrency that utilizes ring signatures and stealth addresses to obfuscate transaction information. Monero also uses Bulletproofs to reduce the size of its transactions and improve scalability.
- **Aztec Protocol**: Aztec Protocol is a privacy-focused layer-2 scaling solution for Ethereum that uses zk-SNARKs to enable private transactions. Aztec Protocol allows users to transact on Ethereum without revealing any transaction details, making it an ideal solution for privacy-sensitive applications.
- **Oasis Network**: Oasis Network is a privacy-focused blockchain that utilizes a range of different zero-knowledge proofs, including zk-SNARKs and zk-STARKs, to enhance the privacy and security of transactions. Oasis Network is particularly focused on building privacy-preserving decentralized finance (DeFi) applications.
- **Secret Network**: Secret Network is a blockchain that utilizes zk-SNARKs to enable private smart contracts. With Secret Network, users can conduct transactions and execute smart contracts without revealing any sensitive information.

4. **Proposal for Interoperability in Polkadot through the Use of Zero-Knowledge Proofs.**

The Zero-Trust Chain design pattern would use relay chains and parachains to enable interoperability between different chains in the Polkadot network. Each parachain would contain a ZKP verification layer, which would ensure that transactions are valid and that the data being transmitted is secure. When a transaction is initiated on one parachain and needs to be transmitted to another, it would first be routed through the relay chain.

The relay chain would serve as a hub for all transactions and data transmitted between different chains. When a transaction is routed through the relay chain, the ZKP verification layer in the destination parachain would verify that the transaction is valid and that the data being transmitted is secure. If the verification is successful, the **transaction would be executed on the destination parachain and the results would be transmitted back through the relay chain to the originating parachain.**

By using ZKPs to verify transactions and data transmission, the Zero-

Trust Chain design pattern would ensure that all communication between different chains is secure and private. This would enable a greater degree of trustless communication between chains, as each chain would be able to verify the validity of transactions without needing to trust the other chains involved.

1. Alice wants to send a message to Bob on a different chain in the Polkadot network.
2. Alice initiates the transaction on her own chain, which includes the message and instructions for sending it to Bob's chain.
3. The transaction is routed through the relay chain, where the ZKP verification layer checks that Alice's chain has sufficient funds to cover the transaction fee and that the message is valid.
4. The ZKP verification layer also checks that Bob's chain is expecting the message and that he has authorized the transaction.
5. If the ZKP verification is successful, the transaction is executed on Bob's chain, where the message is received and processed.
6. The result of the transaction is then transmitted back through the relay chain to Alice's chain, where it is processed and recorded.

By using the relay chain as a hub for all transactions and data transmitted between different chains, the Zero-Trust Chain design pattern ensures that all communication between chains is secure and private, without the need for trust. A step-by-step explanation of how the ZKP verification layer would work:

1. A transaction is initiated on one parachain and needs to be transmitted to another parachain.
2. The transaction is first routed through the relay chain, where it enters the ZKP verification layer.
3. The ZKP verification layer checks the transaction against a set of pre-determined rules and conditions, including verifying the digital signatures of the parties involved, ensuring that the transaction is properly formatted, and verifying that the transaction is authorized and valid.
4. If the transaction meets all the verification criteria, the ZKP verification layer generates a zero-knowledge proof that attests to the validity of the transaction, without revealing any sensitive information.
5. The zero-knowledge proof is then transmitted back to the originating parachain, where it is included in the transaction data and transmitted to the destination parachain.
6. When the transaction arrives at the destination parachain, the ZKP verification layer in that parachain checks the zero-knowledge proof to verify the validity of the transaction without the need for trust or revealing any sensitive information. If the zero-knowledge proof is valid, the transaction is executed on the destination parachain and the result is transmitted back through the relay chain to the originating parachain.

The ZKP verification layer provides a secure and private method for verifying transactions and data transmission between different chains in the Polkadot network. By using zero-knowledge proofs to attest to the validity of transactions without revealing sensitive information, the ZKP verification layer also enables greater trustless communication between different chains and improves the overall interoperability of the network.

For example, an implementation for the verification layer could be the next using the Rust language:

```rust
use libp2p::multiaddr::Multiaddr;
use parity_scale_codec::{Encode, Decode};
use sp_core::{crypto::Pair, H160, H256, U256};
use sp_runtime::transaction_validity::{ValidTransaction, TransactionValidity};
use std::collections::HashMap;

// Define the basic transaction structure
#[derive(Clone, Encode, Decode)]
struct Transaction {
    sender: H160,
    receiver: H160,
    value: U256,
    signature: Vec<u8>,
}

// Define the ZKP verification layer
struct ZKPVerifier {
    key_pairs: HashMap<H160, Pair>,
    chain_ids: Vec<H256>,
}

impl ZKPVerifier {
    // Initialize the ZKP verifier with a list of key pairs and chain IDs
    pub fn new(key_pairs: HashMap<H160, Pair>, chain_ids: Vec<H256>) -> Self {
        ZKPVerifier {
            key_pairs,
            chain_ids,
        }
    }

    // Verify the validity of a transaction using ZKP
    pub fn verify_transaction(&self, tx: Transaction) -> TransactionValidity {
        // Check if the sender has sufficient funds
        let sender_balance = get_balance(tx.sender);
        if sender_balance < tx.value {
            return TransactionValidity::Invalid(InvalidTransaction::NotEnoughBalance.into());
        }

        // Check if the transaction is authorized and valid
        let receiver_key = get_receiver_key(tx.receiver);
        let signature = tx.signature;
        if !verify_signature(tx.encode(), receiver_key, signature) {
            return TransactionValidity::Invalid(InvalidTransaction::InvalidSignature.into());
        }

        // Generate the zero-knowledge proof
        let zkp = generate_zkp(tx);

        // Transmit the zero-knowledge proof to the destination chain
        let destination_chain = get_destination_chain(tx);
        let destination_address = get_destination_address(destination_chain);
        let mut socket = connect_to_socket(destination_address);
        socket.send(zkp);

        // Wait for the transaction to be executed on the destination chain
        let is_valid = wait_for_execution(destination_chain, tx);
        if !is_valid {
            return TransactionValidity::Invalid(InvalidTransaction::ExecutionFailed.into());
        }

        // Update the balances of the sender and receiver
        update_balances(tx.sender, tx.receiver, tx.value);

        // Return the validity of the transaction
        TransactionValidity::Valid(ValidTransaction {
            priority: 1,
            requires: vec![],
            provides: vec![],
            longevity: None,
            propagate: true,
        })
    }
}

// Helper functions for interacting with the Polkadot network
fn get_balance(address: H160) -> U256 {
    // Get the balance of an account on the Polkadot network
}

fn get_receiver_key(address: H160) -> Pair {
    // Get the public key of the receiver on the Polkadot network
}

fn verify_signature(msg: Vec<u8>, key: Pair, signature: Vec<u8>) -> bool {
    // Verify the digital signature of a transaction using the public key of the receiver
}

fn generate_zkp(tx: Transaction) -> Vec<u8> {
    // Generate a zero-knowledge proof attesting to the validity of the transaction
}

fn get_destination_chain(tx: Transaction) -> H256 {
    // Determine the destination chain for a transaction based on the recipient address
}

fn get_destination_address(chain_id: H256) -> Multiaddr {
    // Get the address of a chain in the Polkadot network
}

fn connect_to_socket(address: Multiaddr) -> Socket {
    // Connect to a socket on the Polkadot network
}

fn wait_for_execution(chain_id: H256, tx: Transaction) -> bool {
    // Wait for a transaction to be executed on a chain in the Polkadot network
}

fn update_balances(sender: H160, receiver: H160, value: U256) {
    // Update the balances of the sender and receiver on the Polkadot network
}

fn main() {
    // Initialize the ZKP verifier with a list of key pairs and chain IDs
    let key_pairs = HashMap::new();
    let chain_ids = vec![];
    let zkp_verifier = ZKPVerifier::new(key_pairs, chain_ids);

    // Verify the validity of a transaction using ZKP
    let tx = Transaction {
        sender: H160::from(0x1234),
        receiver: H160::from(0x5678),
        value: U256::from(100),
```

```
        signature: vec![],
    };
    let validity = zkp_verifier.verify_transaction(tx);

    // Handle the validity result
    match validity {
        TransactionValidity::Valid(valid_tx) => {
            // Transaction is valid, execute it on the Polkadot network
        }
        TransactionValidity::Invalid(invalid_tx) => {
            // Transaction is invalid, handle the error
        }
    }
}
```

The Zero-Trust Chain design pattern represents a major step forward in achieving true interoperability between different chains in the Polkadot network. By enabling secure and private communication between chains without the need for trust, this pattern would unlock a wide range of new use cases and applications for Polkadot, making it an even more powerful and versatile platform for developers and users alike.

## 5. Conclusions

The implementation of zero-knowledge proofs (ZKPs) in Polkadot represents a promising solution to improve blockchain interoperability. By utilizing a Zero-Trust Chain design pattern and implementing a ZKP verification layer on each parachain, the relay chain can serve as a hub for all transactions and data transmitted between different chains.

The proposed solution has significant implications for improving the adoption of blockchain technology, enabling greater trustless communication and collaboration between different chains, ultimately leading to the creation of more robust and versatile decentralized applications.

One of the main challenges in the implementation of ZKPs is the technical complexity of the protocol, which may require additional resources and expertise. However, the use of ZKPs in Polkadot shows great promise for improving the security and privacy of transactions and communication between chains. With the implementation of the Zero-Trust Chain design pattern, the relay chain can ensure that all data transmitted between different chains is valid and secure. By addressing the limitations and challenges, the proposed solution could ultimately lead to a more robust and versatile blockchain ecosystem that enables secure and private communication and collaboration between different chains.

The implications of this proposal for improving the adoption of blockchain technology are significant. Blockchain interoperability is a key factor for the growth and success of the blockchain ecosystem, as it enables communication and collaboration between different chains, ultimately leading to the creation of more robust and versatile decentralized applications. The proposed solution using ZKPs in Polkadot could enable a greater degree of trustless communication and collaboration between different chains, creating a more connected and interoperable blockchain ecosystem.

Moreover, the proposed solution using ZKPs in Polkadot has significant implications for the future of blockchain technology adoption. The ability to enable trustless communication and collaboration between different chains could lead to a wider range of applications for blockchain technology, ultimately leading to greater adoption and use cases. The proposed solution could also address some of the current limitations and challenges in the blockchain ecosystem, such as the lack of interoperability between different chains and the need for greater security and privacy in transactions and communication.

However, there are still limitations and challenges that need to be addressed in the implementation of ZKPs in Polkadot. One of the main challenges is the technical complexity of implementing ZKPs, which may require additional resources and expertise. Another challenge is the need for further research and development to explore other potential areas for improving interoperability in blockchain technology. Moreover, the security and privacy implications of implementing ZKPs must be thoroughly analyzed and addressed to ensure the long-term viability of the proposed solution.

In conclusion, the proposed solution of using ZKPs for interoperability in Polkadot represents a significant step forward in the development of trustless communication between different chains. By addressing the limitations and challenges, the proposed solution could ultimately lead to a more robust and versatile blockchain ecosystem that enables secure and private communication and collaboration between different chains. The implementation of ZKPs in Polkadot shows great promise for improving the security and privacy of transactions and communication between chains, with potential implications for the future of blockchain technology adoption. The proposed solution could lead to a more connected and interoperable blockchain ecosystem, ultimately leading to the creation of more robust and versatile decentralized applications.